\documentclass[a4paper,11pt]{article}

\usepackage{jcappub}

\usepackage{amsmath}
\usepackage{amssymb}
\usepackage{amsfonts}
\usepackage{mathtools}

\usepackage{lineno}
\usepackage{lscape}
\usepackage{soul}
\usepackage{subfigure}
\usepackage{graphicx}
\usepackage{color}
\usepackage{bm}
\usepackage{epstopdf}
\usepackage{url}
\usepackage{booktabs}
\usepackage{mdframed}
\usepackage{booktabs}
\usepackage{siunitx}
\usepackage{multirow}
\usepackage[dvipsnames]{xcolor}

\definecolor{burgundy}{cmyk}{0,1,0.75,0.50}

\def\skyfact{\texttt{SkyFACT}\space}
\usepackage[utf8x]{inputenc}
\usepackage[normalem]{ulem}

\newcommand{\sconf}[5]{
	$\left[\begin{smallmatrix}
	{#1} & {#2} & {#3} \\ {#4} & {#5} & \cdot
	\end{smallmatrix}\right]$
}

\begin{document}

\title{
\color{burgundy} 
On the progressive hardening of the cosmic-ray proton spectrum in the inner  Galaxy
}

\author[a]{Mart Pothast}
\emailAdd{mart.pothast@student.uva.nl}
\author[a]{Daniele Gaggero}
\emailAdd{daniele.gaggero@gmail.com}
\author[a]{Emma Storm}
\emailAdd{e.storm@uva.nl}
\author[a]{Christoph Weniger}
\emailAdd{c.weniger@uva.nl}

\affiliation[a]{Gravitation Astroparticle Physics Amsterdam (GRAPPA),\\
Institute for Theoretical Physics Amsterdam and Delta Institute for Theoretical Physics,\\ University of Amsterdam, Science Park 904, 1090 GL Amsterdam, The Netherlands
}

\date{\today}

\begin{abstract}
{
Spatial variations of the average properties that characterize the hadronic component of the diffuse Galactic cosmic-ray sea, in particular the spectral slope and normalization, may unveil critical information about their confinement mechanism in the Galaxy. In the first part of this paper we perform an analysis of the Fermi LAT gamma-ray data with the \skyfact package, which combines image reconstruction techniques with standard template fitting, isolate the hadronic emission and decompose it into Galactocentric rings. We find a significant hardening of the hadronic spectral index towards the molecular ring. We study this hardening in different energy ranges, and assess its resilience with respect to different prescriptions in the analysis setup. In the second part we quantify the contribution to the diffuse gamma-ray flux coming from unresolved point sources with a dedicated Monte Carlo simulation, and consider whether the trend characterized in the first part can be mimicked by a progressively more relevant flux associated to this component in the inner Galaxy. We find that the observed hardening of the hadronic spectral index cannot be due to unresolved sources in the sub-TeV energy range, especially outside the molecular ring, given reasonable assumptions about the unresolved source population.
}
\end{abstract}

\keywords{Galactic cosmic rays, gamma ray astronomy, statistical analysis}

\maketitle


\section{Introduction}

The study of the gamma-ray emission from the Galactic plane provides a unique opportunity to understand the transport properties of cosmic rays (CRs) in different regions of the Galaxy. The basic picture for the production and propagation of Galactic CRs, which dates back to pioneering studies in the early 1960s \cite{ginzburg1964}, is based on the presence of a diffuse {\it sea} of high-energy particles confined in a large {\it diffusive halo}. The motion of those particles is described by a random walk, relatively easily modeled in terms of a homogeneous, isotropic diffusion equation, with the addition of advective and loss terms. Although the actual physical processes characterizing the interaction between the CR sea and the magnetized, turbulent interstellar plasma, which are supposed to be responsible for their random walk, and ultimately for their confinement, are far from being understood, the phenomenological picture stated above was considered adequate to reproduce the available data for a long time.

Currently, given the high statistics of the most recent CR measurements, and the sharper view of the gamma-ray sky in the multi-GeV domain provided by the Fermi-LAT instrument, it is possible to challenge this standard picture. It is therefore important to identify clear trends in the data, for example associated to spectral changes across the Galaxy, since these features are likely to provide useful insights on the physics of CR transport across the Galaxy.

A remarkable example of a large-scale trend with possible relevant implications for the microphysics of CR transport is the progressive hardening towards the inner Galaxy of the proton slope inferred by gamma-ray data. This feature was first outlined in \cite{Gaggero:2014xla}, and later confirmed by two different model-independent analyses \cite{Acero2016ApJS,Yang:2016jda}.

The first question that naturally arises from those results is whether the trend is actually related to the diffuse component of the gamma-ray sky, and can therefore be the signature of some anomalous or non-standard CR transport process at work, or instead can be ascribed to a progressively larger number of unresolved sources in the inner regions of the Galaxy.

Once this aspect is clarified, it is important to understand possible physical scenarios able to reproduce such a hardening. In this context,
two different explanations were proposed in the literature. 

The scenario presented in \cite{Cerri:2017joy} is based on two pieces of information: the presence of a poloidal component of the regular Galactic magnetic field which is directed in the perpendicular direction with respect to the Galactic plane in the inner Galaxy; and the harder scaling of the parallel (w.r.t. the regular field) diffusion coefficient with rigidity that can be inferred from numerical simulations \cite{DeMarco2007a,Snodin:2015fza}. Given these inputs, the authors build a model based on anisotropic CR transport featuring a progressive transition from perpendicular to parallel vertical escape from the Galaxy, and demonstrate that the picture is compatible with the data. 

In \cite{Recchia2016}, instead, the role of CR self-confinement via the {\it streaming instability} is taken into account. GeV CRs are expected to trigger the growth of the Alfv\'en waves that are in turn responsible for their own confinement, and this effect is expected to be stronger where stronger CR gradients exist, i.e. in the inner Galaxy. As a consequence, the diffusion coefficient in the inner regions of the Galaxy should be much lower (reflecting the more effective confinement), so advection should take over up to large energies and the propagated spectrum should be therefore closer to the injected one, explaining the hardening.

The two interpretations predict different behavior at high energies: In fact, the former is expected to hold at all energies, while the latter is based on a physical effect that becomes progressively less relevant with increasing energy. The high-energy CR flux steeply declines with rigidity. Therefore, at energies $\gtrsim 100$ GeV, their ability to trigger the streaming instability should be negligible \cite{Farmer:2003mz,Blasi:2012yr} \footnote{Actually, as noticed in \cite{Blasi:2012yr}, the transition between propagation in self-generated turbulence and the one in pre-existing turbulence could be responsible for the spectral feature detected by PAMELA and AMS in primary and secondary species at a rigidity of $\simeq 200$ GV}. 

Given this important distinction, a natural way to disentangle the two interpretations is to analyze the high-energy gamma-ray data (above $\simeq 30$ GeV), and identify whether the same trend pointing towards a hardening in the inner Galaxy is present. If so, an interpretation in terms of transport properties, such as anisotropic diffusion, that hold at all energies would be favored.

This paper addresses the two important questions outlined above. In the first part we adopt, for the first time in this context, the {\tt SkyFACT} package \cite{Storm:2017arh}, a flexible statistical tool designed to model and decompose the gamma-ray sky into its various components.  It allows to account for a very large number ($\sim 10^5$) of nuisance parameters in order to marginalize over the imperfections of current models built with CR transport codes.  We use this analysis tool to robustly characterize the dependence of the proton spectral index as a function of the Galactocentric radius {\it in different energy ranges}, with the ultimate goal to shed light on the origin of this anomaly. In particular, we perform for the first time an energy-dependent analysis in order to assess the presence of such a trend in Fermi-LAT data above $30$ GeV, and discuss different systematic effects that can affect the result, including the degeneracy with Inverse Compton emission.

In the second part of the paper, we quantify the relevance of unresolved point sources (UPS) via a population study based on currently available resolved point source catalogs and our current knowledge on the luminosity functions of the most relevant classes of sources, and discuss whether a significant contribution from UPS can mimic the progressive hardening.

\section{Hadronic gamma-ray slope analysis: Method}
\label{sec:proton_slope}

The gamma-ray emission from the Galaxy is due to a variety of mechanisms. CR nuclei and leptons, during their random walk through the interstellar medium (ISM), interact with interstellar gas, magnetic fields and the diffuse low-energy interstellar radiation field (ISRF) emitted by stars and further reprocessed by dust grains. As a result, hadrons produce neutral pions which in turn decay into energetic gamma rays; leptons, on the other hand, are able to up-scatter the ISRF photons, and emit bremsstrahlung radiation as well due to the interaction with the electric fields associated to the ionized component of the ISM. 

In order to gain information on the hadronic CR component and its spectral properties, it is therefore crucial to disentangle these different types of emission, associated to different targets and characterized by different morphologies. To this aim, we perform a novel analysis of the gamma-ray emission from the Galactic plane by exploiting the recently developed \texttt{SkyFACT} tool~\cite{Storm:2017arh}. 

\subsection{\texttt{SkyFACT}}

\texttt{SkyFACT} uses a combination of template fitting and image reconstruction methods to decompose the gamma-ray sky into its various components.  It can account for the (still numerous) imperfections of diffuse emission models by giving the individual model components some flexibility to better match the data.  We give a brief overview of the procedure of fitting the gamma-ray sky using \skyfact but refer the reader to \cite{Storm:2017arh} for a detailed description.

Traditional analysis of diffuse gamma-ray emission relies on fitting gamma-ray data with a linear combination of templates that capture the different emission processes mentioned above. These templates, however, come with significant systematic uncertainties that are often difficult to quantify, which create relevant biases when interpreting the gamma-ray sky. 
\skyfact moves beyond the ``static'' template fitting paradigm and allows the templates to vary in both morphology and spectrum.  Even if the deviation from the original templates remain small, this requires a large number of extra parameters (as many as the number of pixels plus the number of energy bins), which are treated as regularized nuisance parameters.  Regularization terms in the model likelihood are then implemented such that deviations from the original templates get penalized in a way that reflects the expected systematic uncertainties. \skyfact also allows for the possibility to leave complete spectral or spatial freedom for some of the templates, in which case it effectively behaves like an image reconstruction tool for components that cannot be modeled \textit{a priori}.

In more detail, the diffuse gamma-ray photon flux in pixel $p$ and energy bin $b$ is modeled as:
\begin{align}
\label{eq:skyfact_model}
\phi_{pb}=\sum_{k} T^{(k)}_p \tau^{(k)}_p \cdot S^{(k)}_b \sigma^{(k)}_b\cdot \nu^{(k)},
\end{align}
where the sum is over $k$ model emission components and $T$ and $S$ are, respectively, the spatial and spectral templates. The set of parameters $\tau^{(k)}$ and $\sigma^{(k)}$ are spatial and spectral nuisance parameters, respectively, and $\nu^{(k)}$ is the overall normalization parameter, for model component $k$. 
These parameters essentially represent the normalizations per spatial pixel, per energy bin, and overall, for each component. 
The model flux in \autoref{eq:skyfact_model} is then convolved with the point spread function and exposure to give the expected number of photons in each bin.

As mentioned above, in order to constrain the variability of the nuisance parameters $\tau$, $\sigma$, and $\nu$, additional regularization terms are added to the likelihood. These terms are controlled by three hyper-parameters ($\lambda$, $\lambda'$ and $\lambda''$) that constrain the spatial, spectral and overall modulation. 
We adopt the maximum entropy method (MEM) as a regularization algorithm, a technique commonly used in the context of astronomical image reconstruction \cite{Skilling1979}. This method is designed in order not to introduce any spurious structures or patterns in the data. Hence, the reconstructed images we get appear as featureless as possible, while still being consistent with the input data.

\skyfact adopts the {\tt L-BFGS-B} \cite{LBFGSB} algorithm to perform maximum likelihood estimation. The final likelihood takes the following form: 
\begin{equation}
\ln\mathcal{L} = \ln \mathcal{L}_P + \ln \mathcal{L}_R \;\;,
\end{equation}
where $\mathcal{L}_P$ is a standard Poisson likelihood and $\mathcal{L}_R$ represents the set of regularization terms (MEM) that constrain the nuisance parameters $\tau^{(k)}$, $\sigma^{(k)}$, and $\nu^{(k)}$, with the beforehand set  regularization hyper-parameters. 

We also include some additional smoothing conditions on the spatial and spectral modulation parameters that control the pixel-to-pixel and bin-by-bin variation; this introduces an additional pair of hyper-parameters per component (see \cite{Storm:2017arh} for further details).
Point sources have no spatial modulation or smoothing, so there are 3 more hyper-parameters that control the point source spectral modulation, spectral smoothing, and overall normalization. 

In this way, the strength of the allowed modulation can be tuned for each model component, effectively allowing us to include the uncertainties intrinsic to the model. 
A table of the hyper-parameter values for each model component is shown in \autoref{tab:comp}. 

The statistical uncertainties on the modulation parameters are estimated from the Fisher information matrix, the inverse of which is a lower limit on the covariance \cite{Cramer1999}. It is computationally difficult to directly perform this inversion, so in practice we use Cholesky decomposition to efficiently sample from a multivariate Gaussian such that the covariance of the sampled model parameters is equivalent to the inverse of the Fisher matrix 
(see \cite{Storm:2017arh} for more details). This procedure yields uncertainties that should be interpreted as conservative estimates on the true statistical uncertainties.

\subsection{Fermi-LAT data}

For this study we use 9.3 years, from 4 August 2008 to 6 November 2017, of Fermi-LAT Pass 8 data. We select only \verb|ULTRACLEAN| events (\verb|evclass=512|) and \verb|FRONT| and \verb|BACK| (\verb|evtype=3|) events. A zenith angle cut at $90^{\mathrm{\circ}}$ was used to reduce photons from the Earth's limb and the data quality cuts recommended by the Fermi-LAT collaboration \path{(DATA_QUAL>0) && (LAT_CONFIG==1)} were applied. We use standard LAT data analysis tools as available from the NASA website\footnote{\url{https://fermi.gsfc.nasa.gov/ssc/data/analysis/}} to create counts and exposure maps of $720\times81$ pixels with a pixel size of $0.5$ degrees centered at the Galactic center. The photon energy was binned in 25 log-spaced energy bins between $0.34\text{--}228.65$ GeV and our region of interest spans the whole disk ($|l| < 180^{\mathrm{\circ}}$) with latitudes $|b|<20.25^{\circ}$.

\subsection{Model components}

\begin{figure}[h!]
\includegraphics[width=\textwidth]{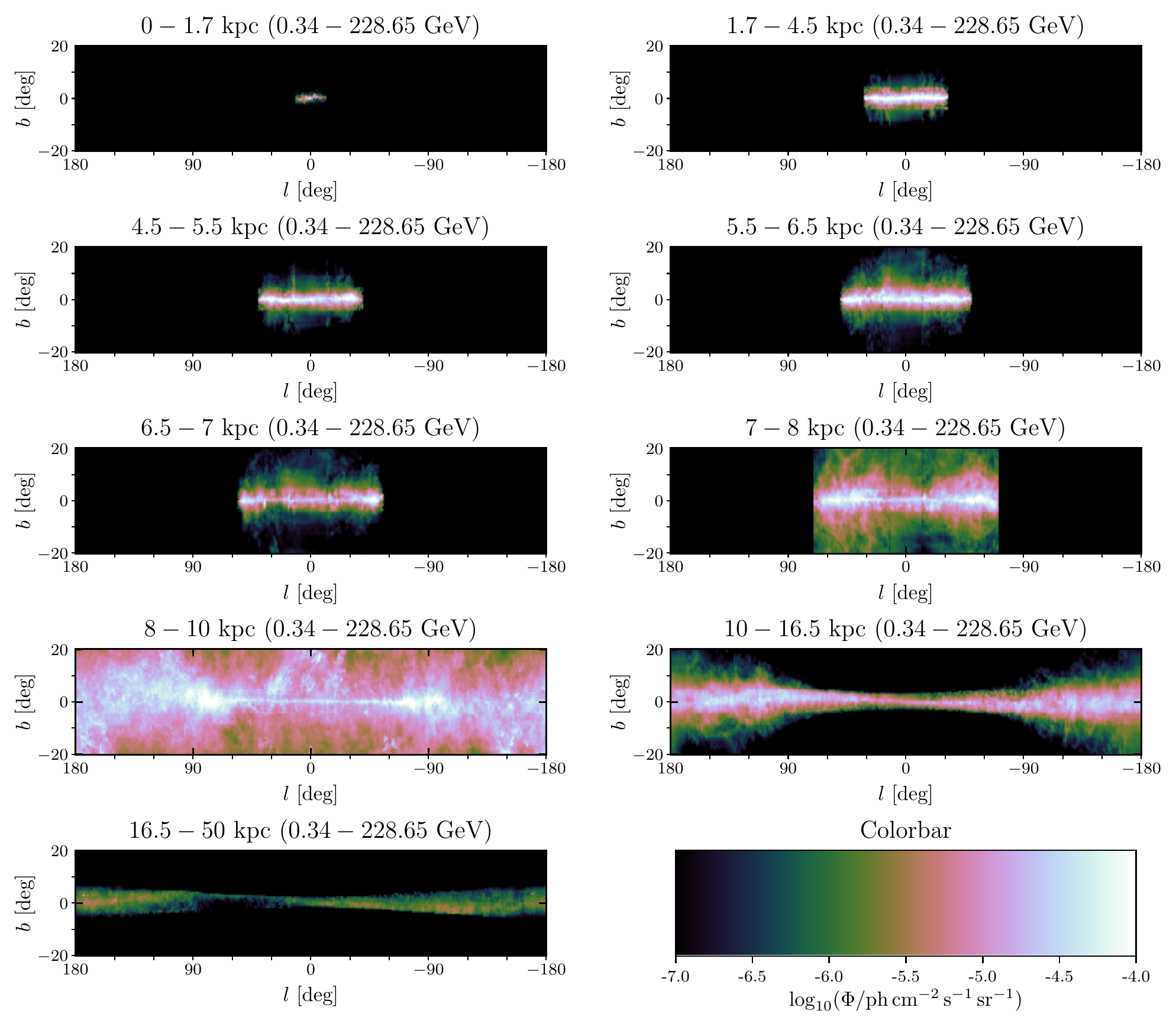}
\caption{{\bf Hadronic emission templates  for different Galactocentric rings}, integrated over all energies. The aspect ratio is altered to increase visibility.}
\label{fig:GR_maps}
\end{figure}

The model components and regularization parameters used in this study are very similar to the settings of the model labeled \texttt{RUN4} in \cite{Storm:2017arh} (see Table 1 therein). We will summarize the most important components here and note important differences between the analysis in this study and the one in \cite{Storm:2017arh}.

We use the gas column densities (HI+CO) as available from the \texttt{GALPROP} public release to build the $\pi^0$ template and bin them in 9 Galactocentric annuli. This binning facilitates comparison with previous work \cite{Acero2016ApJS}. The spin temperature for the HI maps is fixed to $T_S=125 \; \rm K$ and the molecular hydrogen proportionality constant $X_{\rm CO}=1.9 \times 10^{20} \rm \ cm^{-2}/(K \ km/s)$. 

The initial spectrum for the $\pi^0$ templates is taken from Ref. \cite{Ackermann2012}. This spectrum is in agreement with local cosmic-ray observables and approximately follows a power-law scaling with an index of 2.7 at energies $> 2$ GeV. The spectral regularization parameter ($\lambda'$) of the gas rings is chosen so that the spectra are constrained within 25\% of the initial spectrum. The spatial regularization is set so that variations are within $\sim 32 \%$ of the initial template. The best-fit gas templates are shown in \autoref{fig:GR_maps}.

Note that we do not implement any correction to take into account dark gas, i.e. gas that is not resolved by the 21cm or CO line emission, because it was shown that this can be recovered through the use of the modulation parameters \cite{Storm:2017arh}. 

The ICS model we used was created with \texttt{DRAGON} and \texttt{GammaSky} \cite{Evoli:2008dv,Bernardo2013, Evoli2012a}. We use the ISRF as documented in \cite{Porter:2005qx} with the Ferri\`ere source model \cite{Ferriere2001} and the `KRA4' cosmic-ray propagation model from \cite{2013JCAP...03..036D}. The initial ICS spectrum is again taken from \cite{Ackermann2012} and allowed to vary by 25\%. The ICS spatial template is constrained to within a factor two.

A notable difference between this work and \cite{Storm:2017arh} is that we do not add a template for the GeV excess in the Galactic center and use a more conservative template for the \textit{Fermi} bubbles, following the region as defined in \cite{su2010}, and allow no spatial variation for this bubble template. The spectrum of the bubbles is taken from \cite{Ackermann2014} and is constrained to within 1\%. We explore the effects of these choices on the final results in \autoref{sec:sys}. Furthermore, we add the necessary point sources and extended sources within our ROI from the {\tt 3FGL} catalog \cite{Acero2015a} and treat them as described in \cite{Storm:2017arh}.

\begin{table}[h]
	\centering
    \setlength{\tabcolsep}{0.5cm}
	\begin{tabular}{lp{7cm}c}
		\toprule
        
		Components       & Notes & \sconf{\lambda}{\lambda'}{\lambda''}{\eta}{\eta'}  \\ [0.2cm]
		\midrule
		IGRB  & Fixed isotropic template, 25\% spectral freedom. 
        &\sconf\infty{16}\infty00 \\[0.2cm]
		3FGL PSC & Fixed positions, 5\% spectral freedom, 30\% freedom on normalizations.
        &\sconf\cdot{25}{10}\cdot0      \\[0.2cm]
		Extended Sources & Free spectra and templates, mild spatial smoothing.
        &\sconf01\infty40         \\[0.2cm]
		Fermi bubbles & Fixed template, 1\% spectral freedom
        &\sconf\infty{10000}\infty00   \\[0.2cm]
		ICS & Factor of 3 spatial freedom, 25\% spectral freedom, strong spatial smoothing.
        &\sconf1{16}0{100}0    \\[0.1cm]
		Gas rings & 30\% spatial freedom, 25\% spectral freeom, mild spatial smoothing.
        &$9 \times$ \sconf{10}{16}0{25}0 	\\[0.2cm]
        \bottomrule
	\end{tabular}
		\caption{An overview of the model components and the regularization hyper-parameters as used in the gamma-ray sky fit with \skyfact. The second column contains a description of the modeling uncertainties that we allow in the fit.  The matrix contains the corresponding regularizing hyper-parameters, where $\lambda$, $\lambda'$ and $\lambda''$ are spatial, spectral and overall modulation parameters respectively and $\eta$ and $\eta'$ the spatial and spectral smoothing.}
		\label{tab:comp}
\end{table}

\section{Hadronic gamma-ray slope analysis: Results}
\label{sec:proton_slope_results}

\begin{figure}[h]
\centering
\includegraphics[width=\textwidth]{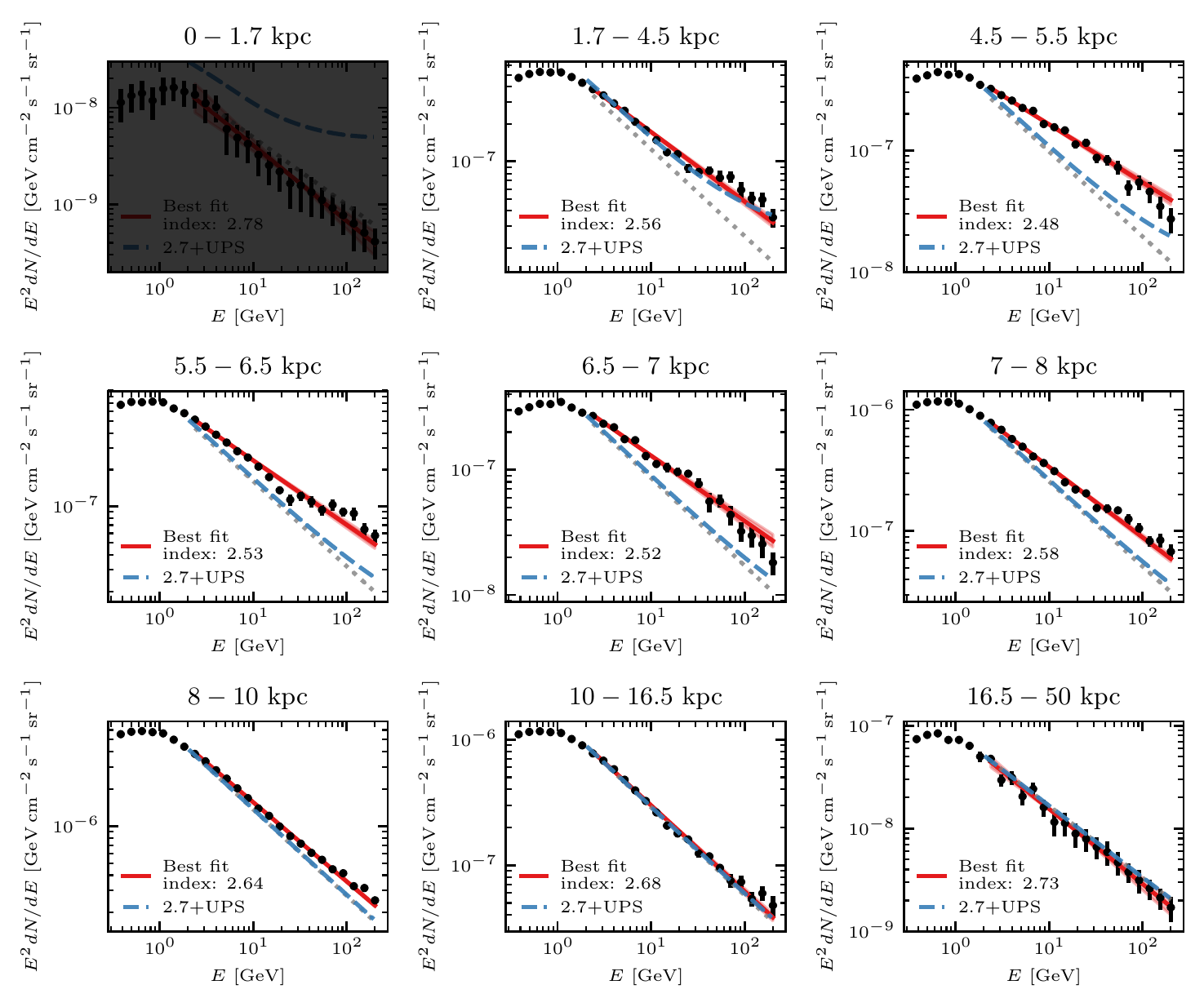}
\caption{{\bf Hadronic gamma-ray spectra in different gas rings}. The black points represent the best-fit spectra to each gas ring template recovered from \skyfact. The red line corresponds to
the best-fit power-law from $2\text{--}228.65\; \rm GeV$, and the pink band is the 68\% credible interval. The gray dotted line shows a power-law with an index of 2.7 for comparison. The blue dashed line shows the contribution of unresolved point source for our reference scenario, added to the reference power-law (see \autoref{sec:UPS}). The first radial bin is grayed out because the fit in that bin is strongly affected by systematic uncertainties (see \autoref{sec:hardening_full_range}). }
\label{fig:spec_GR}
\end{figure}

\begin{figure}[h]
\centering
\includegraphics[width=0.85\textwidth]{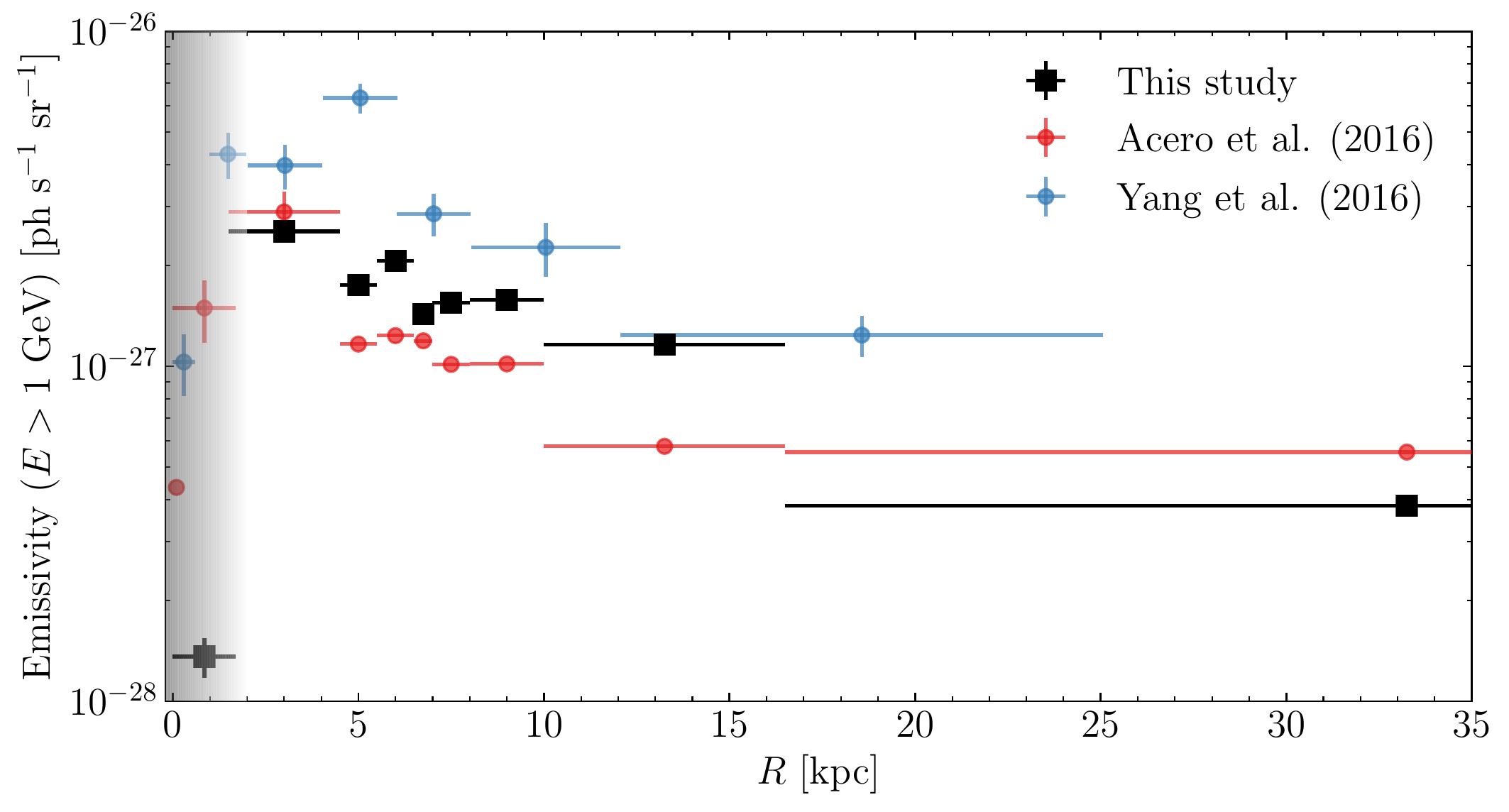}
\caption{{\bf Radial distribution of the gamma-ray emissivity per H atom.} The emissivity is integrated over the $1\text{--}100 \; \rm GeV$ range: This quantity is a proxy to the total CR flux.
Previous studies mentioned in the text are shown for comparison: We notice that Ref. \cite{Acero2016ApJS} provide the emissivity per H atom at $2\; \rm GeV$. The result associated to the first radial bin, corresponding to the inner Galactic bulge, is less reliable for several reasons discussed in the text, and is therefore grayed out in the plot.}
\label{fig:emissivity}
\end{figure}

\begin{figure}[h]
	\centering
	\includegraphics[width=0.85\textwidth]{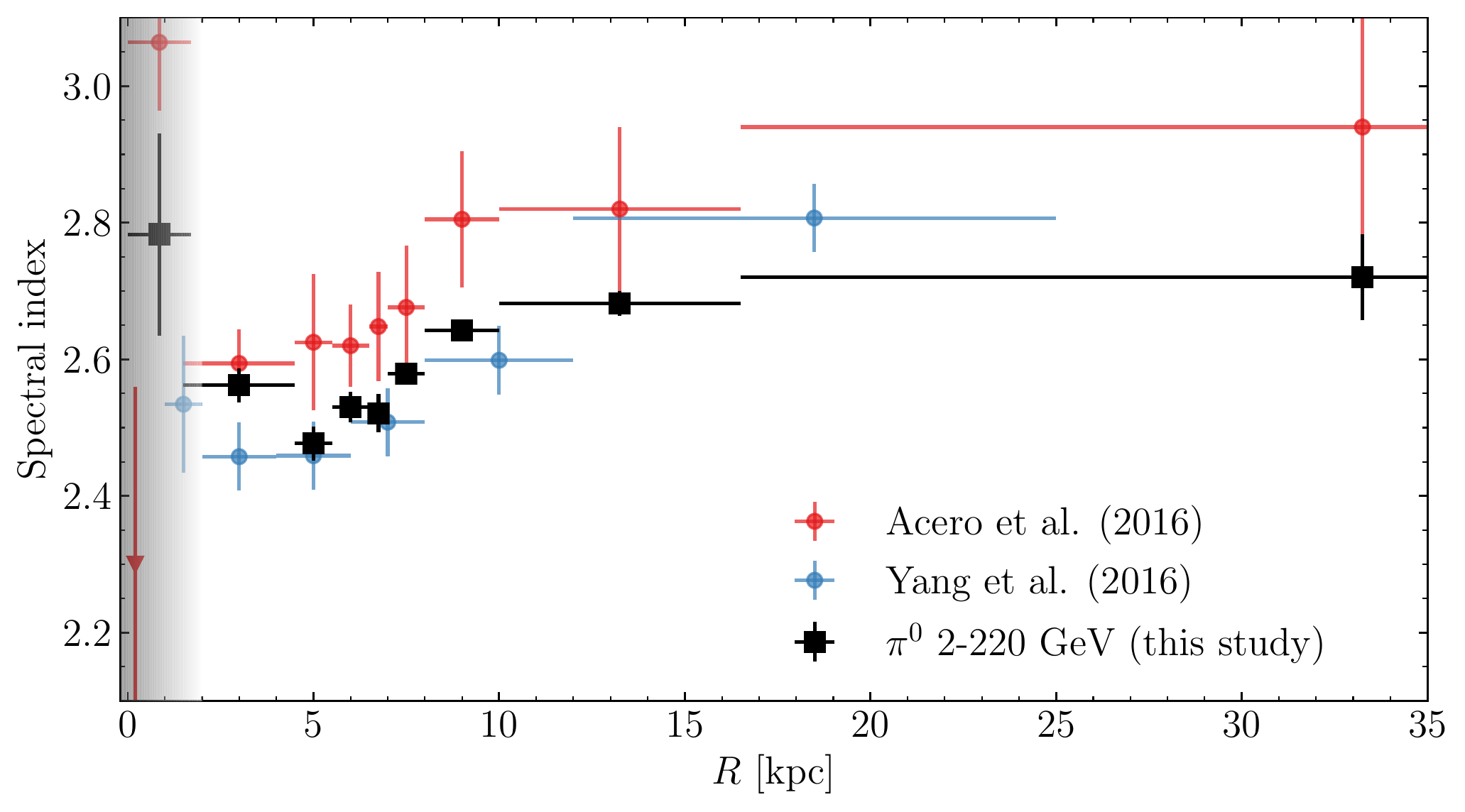}
	\caption{{\bf Spectral index of the hadronic emission for different Galactocentric rings}. We show the spectral index fitted as explained in the text from $2\text{--}220\; \rm GeV$ compared to the trend found in \cite{Acero2016ApJS,Yang:2016jda}. Horizontal error bars indicate bin width in $R$ and vertical error bars are $68\%$ credible intervals.}
	\label{fig:index_allrings}
\end{figure}

\subsection{Hardening of the proton spectrum in the full energy range}
\label{sec:hardening_full_range}



In \autoref{fig:spec_GR}, the best-fit spectra for each Galactocentric annulus from \skyfact\ are shown with black points. In order to identify any trends in the spectra across different rings, we then fit the black points with a single power law in the $2-228.65$~GeV energy range. We are ultimately interested, however, in trends present in the CR proton spectrum. Above a gamma-ray energy $E_\gamma \sim 2$~GeV (the location of the $\pi^0$ bump), the spectrum of the gamma-ray emission from $\pi^0$ decay scales with the underlying CR proton spectrum, with an overall factor of $E_\gamma \approx E_{CR}/10$. Therefore, the photon spectral index of the $\pi^0$-induced gamma-ray emission for $E_\gamma>2$~GeV is directly comparable to the CR proton spectral index.

To perform a power-law fit to the black points above $2$~GeV in \autoref{fig:spec_GR}, we assume the photon flux in each energy bin is normally distributed with a standard deviation as sampled from the inverse Fisher matrix and set up a Markov Chain Monte Carlo (MCMC) using the \texttt{emcee} package \cite{Foreman2013} to sample the posterior distribution of the power-law index. The mean and 68\% credible intervals of the best-fit power-law is shown in \autoref{fig:spec_GR} for every Galactocentric annulus.

A trend of hardening spectra towards the inner Galaxy is already apparent in \autoref{fig:spec_GR}. The black points and the best-fit power-law index (the red line) are clearly harder than the reference power-law with index of $2.7$, which is the observed slope of the local CR emission \cite[e.g.,][]{Casandjian:2015hja,Neronov:2017lqd}. 

The best-fit photon index from \autoref{fig:spec_GR} is plotted versus radial distance from the Galactic center in \autoref{fig:index_allrings} and compared with previous analyses \cite{Yang:2016jda,Acero2016ApJS}. We also show the hadronic gamma-ray emissivity integrated over energies above 1 GeV (for straightforward comparison with \cite{Yang:2016jda}; \cite{Acero2016ApJS} shows the emissivity at 2 GeV), which is a proxy of the hadronic CR flux, in \autoref{fig:emissivity}. Overall, we find a reasonable agreement with both studies. 

As shown in \autoref{fig:emissivity}, the gamma-ray emissivity decreases as a function of the Galactocentric radius in the region beyond the molecular ring, which is located in the second ring corresponding to the $1.7-4.5$ kpc range. The decline in emissivity is not as steep as naively predicted in the context of isotropic and homogeneous diffusive CR transport, as already pointed out several times even before Fermi-LAT data (the so-called {\it gradient problem}: see e.g. \cite{2004A&A...422L..47S} and the more recent discussion in \cite{Evoli:2012ha,Recchia2016}). 

\begin{figure}[h!]
	\centering
	\includegraphics[width=0.85\textwidth]{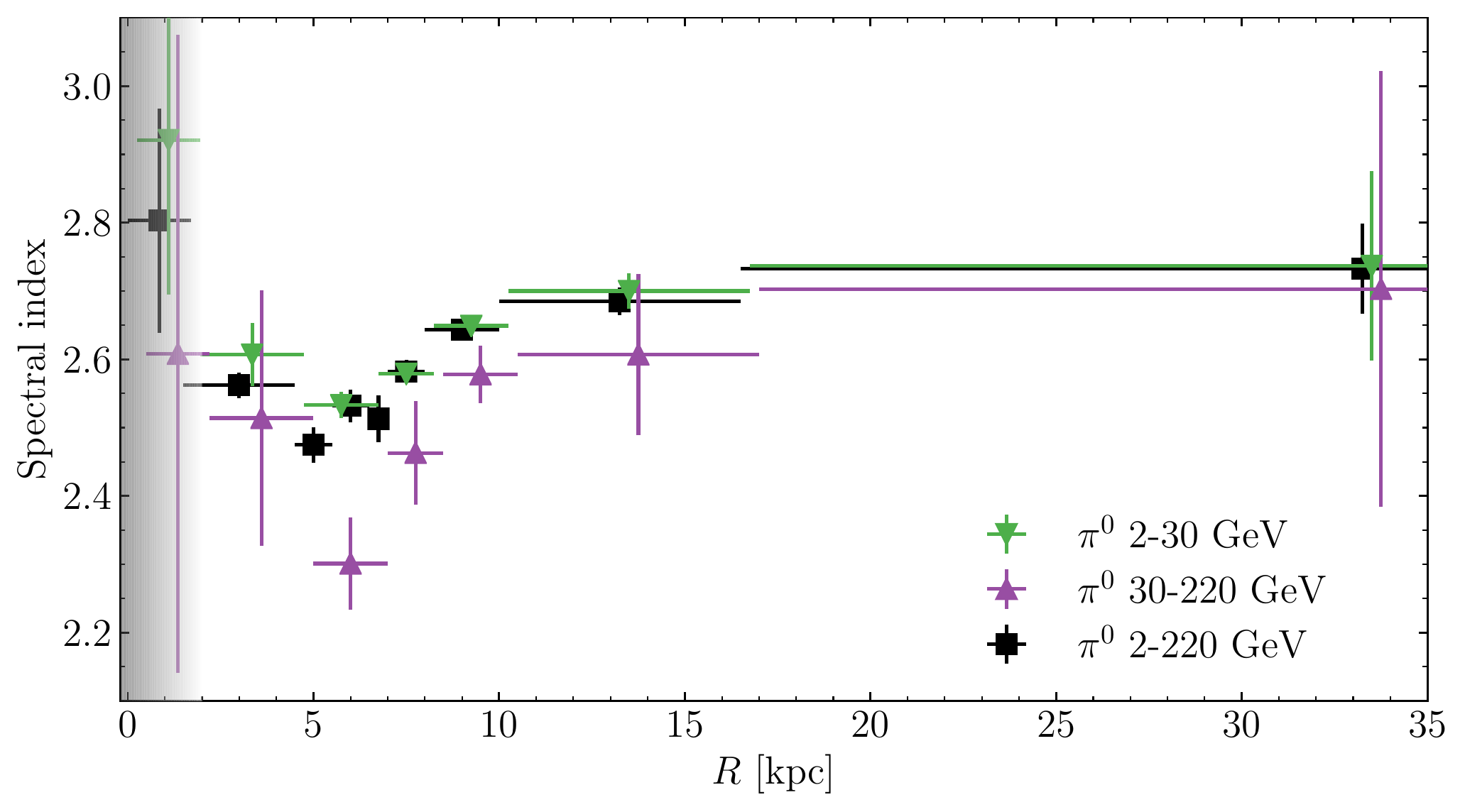}
	\caption{{\bf Energy-dependent analysis of the hadronic gamma-ray slope.} We show the spectral index of gamma rays associated with $\pi^0$ decay at different radii from the Galactic center. The power-law fit is first performed in the full range, and then restricted to both the low-energy (from $2\text{--}30$ GeV), and the high-energy ($30\text{--}220\; \rm GeV$) domain; the low-energy and high-energy points are artificially shifted to the right by 0.25 kpc and 0.5 kpc respectively, so that the error bars do not overlap.}
	\label{fig:index}
\end{figure}

\autoref{fig:index_allrings} clearly shows the progressive hardening towards the Galatic center in a wide range of radii outside the Galactic bulge. 
This trend is qualitatively compatible with previous studies. The uncertainties on the power-law index in this work are smaller than previous analyses due to a combination of different effects, in particular more statistics with respect to previous papers, and the different technique implemented in \skyfact. We discuss these issues in more detail in \autoref{sec:sys}.

The first radial bin from $0\text{--}1.7$~kpc, which contains the Galactic bulge, is quite problematic, and different studies find different results. This is not too surprising, given the small number of pixels associated to this first ring, which results in a low normalization of the hadronic component associated to this bin as a result of our fitting procedure due to lack of constraining power. More importantly, there is large degeneracy between the different components in the Galactic center region, which is by far the most challenging issue both from the observational and modeling point of view. However, a lower-than-average CR flux in that region may be compatible with a scenario characterized by a very fast diffusive escape in the vertical direction, parallel with respect to the direction of the regular magnetic field, if the diffusion tensor is highly anisotropic (see Fig. 4 in \cite{Cerri:2017joy}).

\subsection{Energy dependence of the hardening of the proton spectrum}

We want to assess whether the trend in the spectral hardening is still present at higher energies. This would help to discriminate between the two theoretical interpretations discussed above. In particular, the streaming instability, mostly taking place in the hot ionized phase of the interstellar medium, is not expected to be the dominant confinement mechanism for CR hadrons more energetic than $\sim 100$ GeV, because the CR flux, and therefore the associated pressure, is not large enough to effectively trigger the growth of Alfv\'enic perturbations. As mentioned in the Introduction, this was already noticed in \cite{Farmer:2003mz}, building on previous works (e.g.~\cite{Cesarsky:1980pm,1971ApL.....8..189K}), and recently discussed in \cite{Blasi:2012yr,Evoli:2018nmb} in the context of the local spectral feature in CR nuclei detected by PAMELA and AMS. As a consequence, higher energy CRs should be confined by the interaction with pre-existing turbulence, and the argument presented in \cite{Recchia2016} cannot be invoked. The model based on anisotropic transport, on the other hand, should hold at all energies \cite{Cerri:2017joy}. The assessment of the trend for gamma-ray energies above few tens of GeV (which trace CR hadrons more energetic than $\sim 100$ GeV) is therefore crucial in this context.

In order to search for the presence of a hardening trend at high energies, we model the gamma-ray sky as described above, but due to the increasingly lower photon counts at higher energies, we use a smaller number of gas rings: We merge rings 3 and 4 ($4.5\text{--}6.5\; \rm kpc$), and rings 5 and 6 ($6.5\text{--}8 \; \rm kpc$) and therefore adopt a total of 7 rings instead of 9.
We model the gamma-ray data with \skyfact following the method outlined above; however, we now fit the $\pi^0$ emission with a broken power-law featuring a break at $E_\gamma=30$ GeV.

The spectral index from $2\text{--}30\; \rm GeV$ and from $30\text{--}220 \; \rm GeV$ are shown in \autoref{fig:index} along with our result over all energies from \autoref{fig:index_allrings}. 
The plot shows that the trend is clearly present in the high-energy domain, in particular outside of the Galactic bulge, and is even more pronounced than the one inferred by low-energy gamma rays. 
The reason why the high-energy spectral slopes are systematically harder at all radii, with a more pronounced trend in the molecular ring region, may be due to a combination of different effects. The spectral feature in the CR primary spectrum observed by PAMELA and AMS could be present all through the Galaxy (in particular if interpreted as the transition between the scattering on self-generated and pre-existing turbulence), and is expected to be imprinted in the gamma-ray slope as well. Additionally, unresolved sources are expected to contribute more at larger energies, altering the trend. This effect will be quantified in \autoref{sec:UPS}.

\section{Systematic uncertainties}
\label{sec:sys}

\begin{figure}[h]
\includegraphics[width=\textwidth]{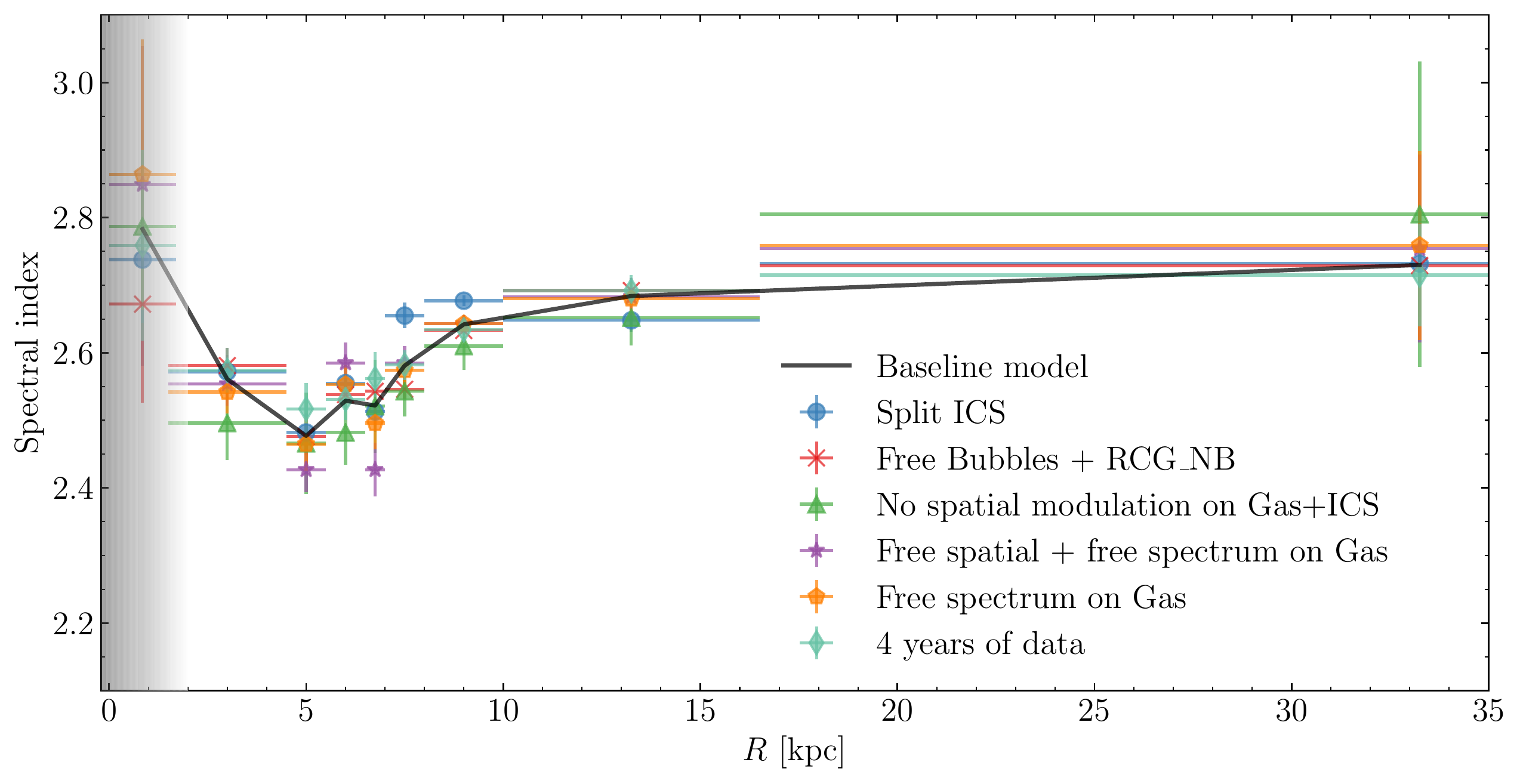}
\caption{{\bf Systematic uncertainties on the hardening.} We show the hadronic spectral indexes in the different Galactocentric bins obtained with \skyfact following different prescriptions, as detailed in the text and summarized in \autoref{tab:sys}.}
\label{fig:systematic}
\end{figure}

Let us now consider the systematic uncertainty associated to our analysis. To this aim, we present several \skyfact analyses performed with different choices in components and modulation parameters. We refer to \autoref{fig:systematic} for a visualization of these results.

First, we consider a fit to the gamma-ray data where the ICS component is split over different radii from the Galactic center: 0--3 kpc, 3--8.3 kpc, 8.3-50 kpc. This splitting is similar to the one applied to the hadronic component, but restricted to three rings only in order to keep the number of fitted parameters under control. We also restrict the ICS spectrum to be more constrained towards the input spectrum. The idea here is that the harder ICS spectrum can pick up some of the residual high energy photons in the inner Galaxy that were previously being absorbed by the gas templates. We see however (blue circles in \autoref{fig:systematic}) that there is still a clear trend of a hardening towards the inner Galaxy.

\begin{table}[b]
	\centering
    \small
    \setlength{\tabcolsep}{0.5cm}
	\begin{tabular}{lp{7cm}}
		\toprule
        Model & Comment \\ \midrule
        Baseline model & As described in \autoref{sec:proton_slope}.\\
        Split ICS & ICS template split over three rings: $0\text{--}3$, $3\text{--}8.3$ and $8.3\text{--}50$ kpc. \\
        Free Bubbles + {\tt RCG\_NB} & Spatial modulation of Fermi Bubbles free to vary with template for the Galactic Center excess as in the best fit from \cite{Bartels2017f}. \\
        No spatial modulation on Gas+ICS & Similar to the traditional analysis as in \cite{Acero2016ApJS,Yang:2016jda}. The spectral modulation is more free: $\lambda'=1$.\\
        Free spectrum on Gas & Spectral modulation on each gas ring is more free: $\lambda'=4$.\\
	Free spatial + free spectrum on Gas & Spectral modulation as above, with more free spatial regularization $\lambda=4$.\\
    4 years of data & Same analysis as \textit{baseline model} but now only with data from the first 4 years of Fermi-LAT. \\
\bottomrule

	\end{tabular}
		\caption{Overview of the several systematics of the gamma-ray fit that were tested in this paper.}
		\label{tab:sys}
\end{table}

We also check whether adding a template for the Galactic center excess and allowing more freedom on the \textit{Fermi} Bubbles affects the trend in the spectral index. In \autoref{fig:systematic} we show (red crosses) the effect of including a Galactic center excess template that follows Red Cluster Giants and the Nuclear Bulge (\texttt{RCG\_NB}), which according to Ref. \cite{Bartels2017f} gives the best fit to the data. 
The addition of these components slightly affects only the first radial bin, but still within the error bars of the \textit{baseline model}.

Then, we perform the analysis with opposite prescriptions regarding the freedom on the different templates: we do not allow any freedom on the spatial modulation and, conversely, more freedom in the spectrum. This setup essentially reproduces the usual rigid template fitting technique adopted in \cite{Acero2016ApJS,Yang:2016jda}. With this prescription, we obtain larger residuals than in the \textit{baseline model} but the trend in the slope is not altered very much (green triangles), except  for the increase of the estimated uncertainties.
We also show that allowing more freedom in the spatial and spectral regularization still qualitatively reproduces the same trend (small purple crosses) -- there seems to be some degeneracy between radial bins when allowing more spatial freedom -- with slightly larger error bars, as expected. The size of the error bars is mostly affected by the spectral regularization so that increasing the spectral variation from 25\% to 50\% leads to a twofold increase of the error bar on the spectral index (orange pentagons).

Finally, we consider again the \textit{baseline model}, but only using 4 years of data, in order to check the impact on the error bars. As expected, the uncertainty increases by the square root of the difference in observation time (cyan diamonds).

\section{Unresolved sources}
\label{sec:UPS}

In order to draw more robust conclusions about the nature of the hardening trend outlined in the previous sections, it is important to quantitatively assess the role of unresolved point sources along the Galactic plane. 
Indeed, a significant population of unresolved hard gamma-ray sources that lie below the detection limit of Fermi-LAT could in principle mimic the progressive hardening towards the inner Galaxy. This would naturally explain why the radial dependence of the proton slope is very similar to the source distribution profile (which peaks at $\simeq 5 \; \rm kpc$), and why the inferred CR spectrum is even harder at higher energies. 
Therefore, in this section we model the expected flux from a population of unresolved hard gamma-ray sources and quantify its relevance with respect to the diffuse $\pi^0$ emission via a dedicated Monte Carlo simulation.

\subsection{Gamma-ray source catalogs}

\begin{figure}[t]
\centering
\includegraphics[width=0.7\textwidth]{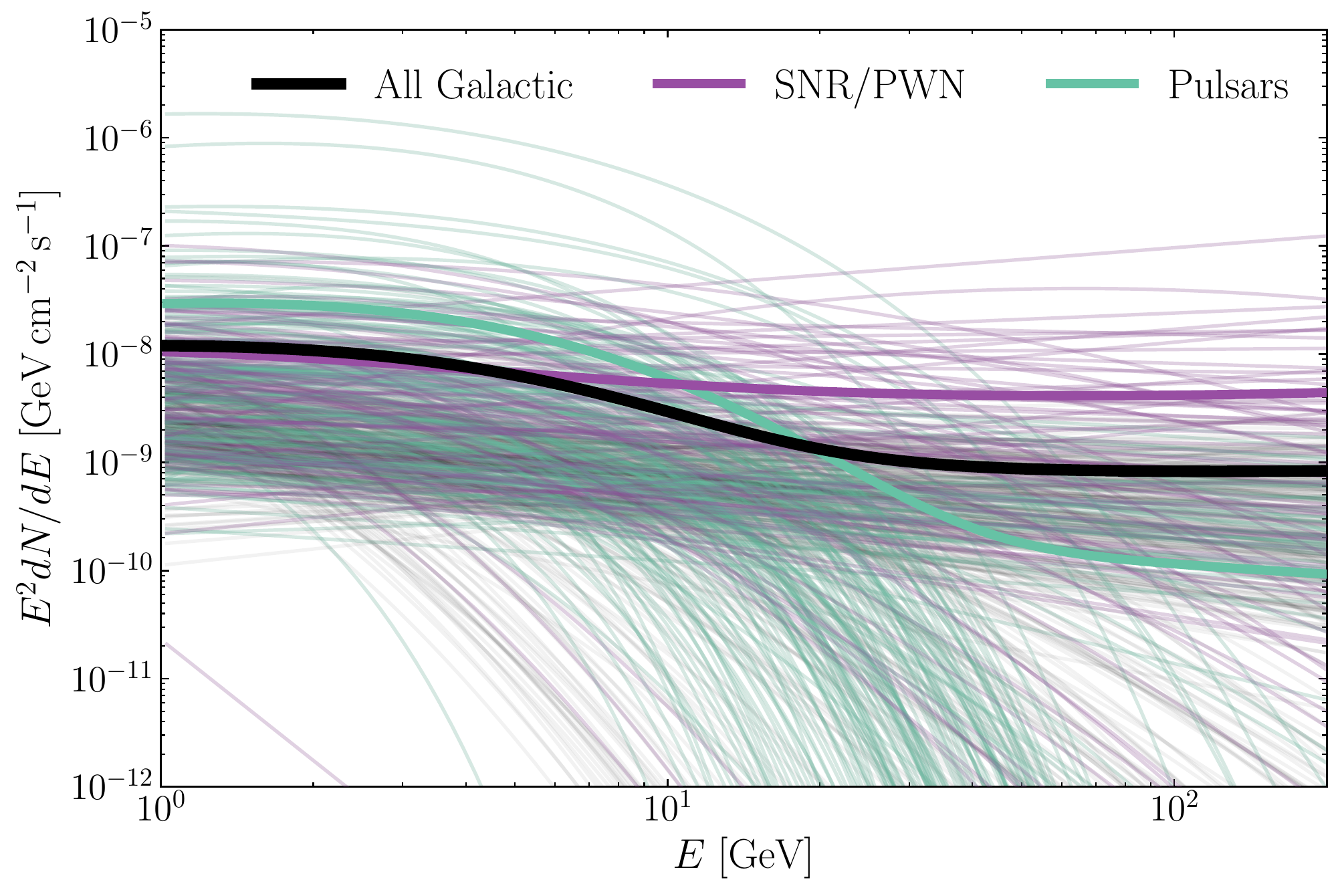}
\caption{{\bf Spectra from various Galactic sources in the {\tt 3FGL} catalog.} 
The thick purple line is the average spectrum of all {\tt 3FGL} sources identified as SNR or PWN. The thin purple lines represent the spectra for all the individual SNR and PWN sources. The thick teal line is the mean spectrum of all {\tt 3FGL} sources identified as pulsars, while the thin teal lines are the spectra of all the individual pulsars. The thick black line is the mean spectrum of all of these sources plus unidentified sources located along the Galatic disk, $|b|<5^{\circ}$.
}
\label{fig:3fgl_spec}
\end{figure}

Population studies of gamma-ray sources in the Galaxy  -- mainly pulsars,  pulsars wind nebulae (PWNe), and shell-type supernova remnants (SNRs) -- are available from the {\tt 3FGL} \cite{Acero2015a} and {\tt 1FHL} \cite{Ackermann2013} catalogs. 
In particular, the latest reference model from the {\tt 3FGL} catalog, based on the spatial distribution of pulsars and on their spectra in the $1\text{--}100$ GeV range, features $2.9 \times 10^4$ Galactic sources drawn from a power-law luminosity function characterized by a slope of $-1.8$ and a luminosity range $2\times 10^{34}\text{--}2\times 10^{39} \; \rm ph \; s^{-1}$. The conclusion of that study is that the unresolved sources in the inner Galaxy below $1 \times 10^{-9} \ \rm ph \ cm^{-2} \ s^{-1} $ amount to 3\% of the total diffuse emission at 1 GeV.

In this section we aim to characterize the relevance of unresolved gamma-ray sources with particular focus on the contribution at higher energies. Given that the average spectrum of point sources is harder than the diffuse emission, the contribution of point sources is expected to increase with energy and can become much larger than the value of 3\% quoted above.

\subsection{Estimating the contribution from unresolved point sources}
\label{back-of-the-envolope}
    
	It is useful to make a back-of-the-envelope estimate of the amount of flux from unresolved sources that one needs to induce a harder observed spectrum. 
    
    Let us assume that the true $\pi^0$ spectrum has a declining slope with index $\gamma$, while the measured $\pi^0$ spectrum has a slope with index $\alpha$ and is contaminated by unresolved sources:
    \begin{equation}
    \Phi_\mathrm{\rm Measured}(E) \simeq \Phi_1\left(\frac{E}{E_0}\right)^{-\alpha} \sim \Phi_\mathrm{UPS}(E) + \Phi_0\left(\frac{E}{E_0}\right)^{-\gamma}\;,
    \end{equation}
where $\Phi_{\rm UPS}(E)$ is the flux from unresolved point sources, and the `$\sim$' indicates that the same measured spectrum can be fitted with different model spectra.  Furthermore, $\Phi_0$ and $\Phi_1$ are free normalization factors, which can in principle be obtained from fits to the data.
    
    The fraction $f$ of the flux from unresolved point sources with respect to the measured $\pi^0$ spectrum can then  be written as follows:
    \begin{equation}
    f = \frac{\Phi_\mathrm{UPS}(E)}{\Phi_\mathrm{\rm Measured}(E)} \,\propto\,\, 1 - \frac{\Phi_0}{\Phi_1} \left(\frac{E}{E_0}\right)^{\alpha-\gamma}\;.
    \label{eq:frac}
    \end{equation}
    
    This estimate holds assuming the true and the measured $\pi^0$ spectrum both give a good fit to the gamma-ray data at $E_0= 1$ GeV so that $\Phi_0\simeq\Phi_1$ (which seems to be the case, see \cite{Ackermann2012}). Setting $\alpha=2.5$ and $\gamma=2.7$ (see \autoref{fig:index_allrings}) we find that, at $10\;(100)$ GeV, $\simeq 40\;(60) \%$ of the measured flux would need to be from unresolved sources to induce a hardening from $2.7$ to $2.5$.  

We can also use the equations above to estimate the contribution of unresolved point sources to the  \textit{additional} hardening at $E_\gamma > 30$~GeV on top of the already-present hardening at $E_\gamma < 30 $~ GeV  -- that is, we compare the purple points to the green points in \autoref{fig:index}. In this case, we set $E_0 = E_{\text{break}} = 30 $~GeV, and assume that that $\Phi_0\simeq\Phi_1$, (where $\Phi_0$ and $\Phi_1$ are now normalizations at 30 GeV). Then, at $E = 100$~GeV, unresolved sources would need to be $20\%$ of the measured flux to induce a change in the power-law index from 2.5 to 2.3. 


Following the same train of thought we can estimate the fraction of the {\it integrated} flux above the break energy $E_b$ one needs to induce a break in the spectrum at $E_b$ with index $\Delta \gamma$.
Let us assume that the diffuse flux is described by a single power law. The integrated flux above $E_b$ from a single power law simply reads:
\begin{equation}
\Psi_0 = \int_{E_b}^{\infty} \Phi_b \left(\frac{E}{E_b}\right)^{-\gamma} dE = \frac{1}{\gamma -1}\Phi_b E_b,
\end{equation}
We assume that the addition of the UPS component induces a break: The integrated flux for a broken power law can be approximated by:
\begin{equation}
\Psi_1 \simeq \int_{E_b}^{\infty} \Phi_b \left(\frac{E}{E_b}\right)^{-\gamma+\Delta \gamma} dE = \frac{1}{\gamma -\Delta \gamma -1}\Phi_b E_b.
\end{equation}
Hence, the integrated flux from UPS above $E_b$ can be estimated as $\Psi_{\rm UPS}\approx \Psi_1-\Psi_0$, and the ratio of this flux with respect to the initial single power law can be written as:
\begin{equation}
F \equiv \frac{\Psi_{\rm UPS}}{\Psi_1} = 1- \frac{\Psi_0}{\Psi_1}=\frac{\Delta \gamma}{\gamma-1}.
\end{equation}
Again, under the assumption that both the single and broken power law give a good fit to the data at $\leq E_b$, as seen in \autoref{fig:index}, we get an expression for the fraction $F$ as a function of the spectral index change $\Delta \gamma$:
\begin{equation}
F = \frac{\Delta \gamma}{(\gamma-1)}\;.
\label{eq:F}
\end{equation}
This implies $F \simeq 13 \%$ for a spectral hardening from a slope of $2.5$ to $2.3$ above $30$ GeV.

In the following section, we describe a more quantitative analysis of the estimated fractional contribution of unresolved point sources with a dedicated simulation.

\subsection{Our simulation of Galactic sources}

Our approach to simulate Galactic sources is based on \cite{Acero2015a,Ackermann2013} (which in turn refer to \cite{strong2007}). The code associated with this simulation can be found \href{https://github.com/martp91/mp}{here}\footnote{\url{https://github.com/martp91/mp}}.

\begin{figure}[t]
\centering
\includegraphics[width=0.7\textwidth]{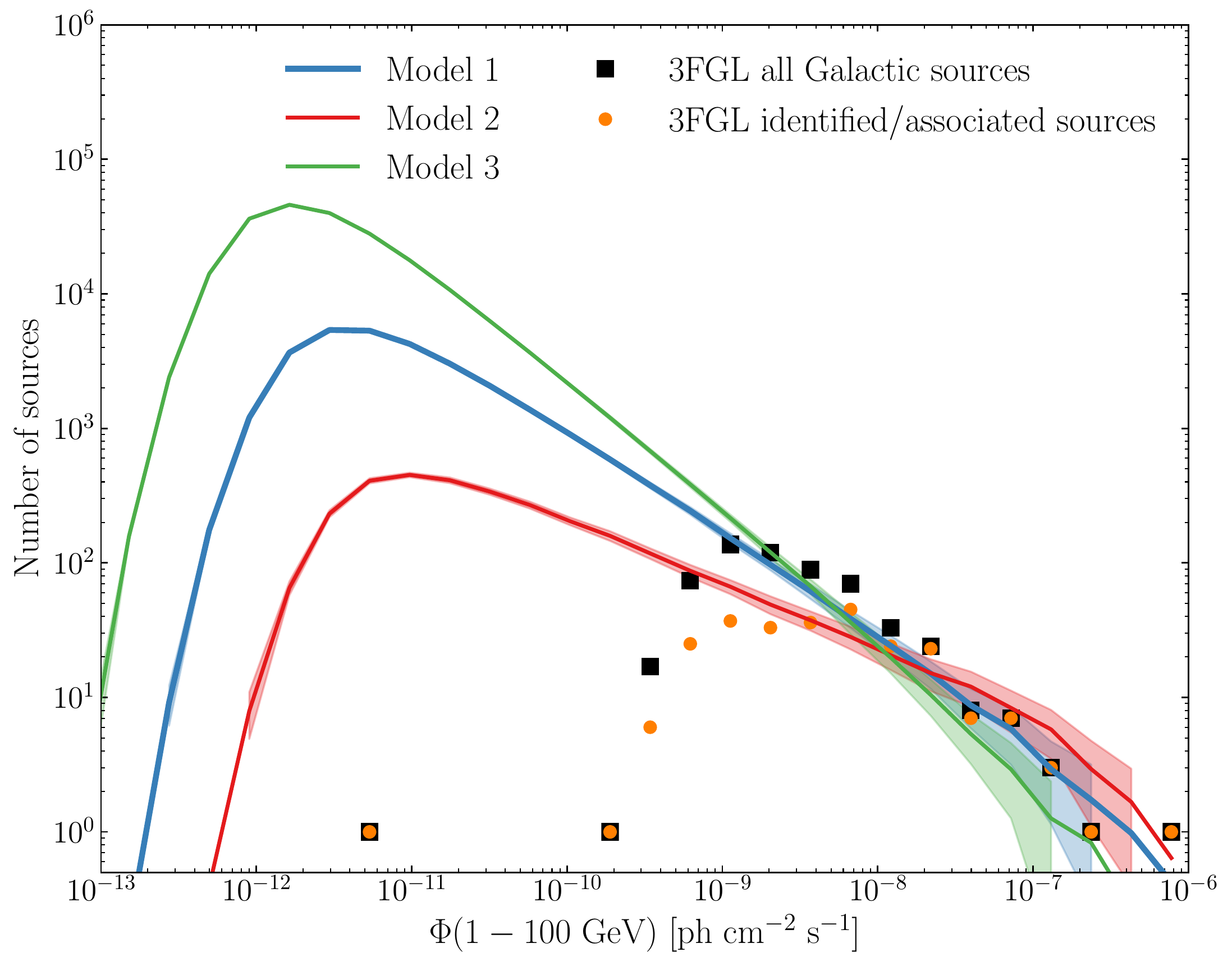}
\caption{{\bf The number of sources per flux interval.} The three models listed in \autoref{tab:lum} are compared with the flux from resolved sources from the {\tt 3FGL} catalog provided by the Fermi-LAT collaboration. The colored bands indicate the estimator of the standard deviation computed on an ensemble of 100 simulations.}
\label{fig:hists}
\end{figure}

The simulation procedure is as follows: We produce ensembles of gamma-ray point sources by drawing the positions from a given spatial distribution in $R$, $\phi$, and $z$. For each source, an integrated luminosity (in the $1\text{--}100 \; \rm GeV$ range) is drawn from a luminosity function.
Then, given distances and luminosities, we calculate a flux for each source. We then assign a random spectrum (in the $1\text{--}200\; \rm GeV$ domain) drawn from fitted spectra in the {\tt 3FGL} catalog from the following classes of sources: associated or identified pulsars (PSRs), supernova remnants (SNRs), pulsar wind nebulae (PWNe), or unidentified sources with latitudes $|b|<5^{\circ}$. The spectra of all considered sources is shown in \autoref{fig:3fgl_spec}. Finally, we tune our result to the actual number of {\it resolved} point sources in the catalog.

Let us now describe the different astrophysical inputs mentioned above in more detail.

The luminosity function (which refers to the $1\text{--}100\; \rm GeV$ range) is a power-law distribution of the form:

\begin{align}
\frac{dN}{dL} = \frac{N(1-a)}{L_{\rm max}^{1-a}-L_{\rm min}^{1-a}} L^{-a},
\end{align}
which features a low-luminosity cutoff $L_{\rm min}$. The integral from $L_{\rm min}$ to $L_{\rm max}$ gives the total number of sources $N$. This simple power-law form for the luminosity function is widely used \cite{Acero:2015prw,Ackermann2013,strong2007,Bartels:2015aea}; in particular, an index $a$ of $\simeq 1.5$ was derived for pulsars in Ref. \cite{strong2007}.

Let us remark that the true form of the luminosity function is uncertain, and the estimated flux from unresolved sources is highly dependent on the exact shape of this ingredient. However, since our only goal is to ascertain whether unresolved sources could potentially mimic a harder spectrum of the diffuse emission and we do not aim at finding the most accurate luminosity function, we adopt the reference model from \cite{Acero2015a} and define two additional scenarios that provide a reasonable bracketing of the uncertainties. A lower limit model can be obtained by choosing a larger value for $L_{\rm min}$, and a smaller (i.e., harder) value of the slope $a$, and \textit{vice versa} for an upper limit model. The parameter $L_{\rm max}$ has a minor impact on our results. We adopt the value $2\times 10^{39} \rm \; ph \;s^{-1}$.

The exact spatial distribution of sources has a percent-level impact on our estimate up to energies of several GeV. Therefore, we only report here  the results obtained from the distribution of pulsars taken from \cite{Lorimer2006}. However, we tested other options \cite{Green2015a,Ferriere2001} as well. The $R$ dependence of the Lorimer type C source distribution can be written as:

\begin{align}
\rho(R) \propto \left(\frac{R}{R_{\odot}}\right)^\alpha \exp{\left(-\beta \frac{R-R_{\odot}}{R_{\odot}}\right)}
\end{align}
and the $z$ dependence is 
\begin{align}
\rho(z) \propto \exp{\left(-|z|/H\right)},
\end{align}
with $\alpha=1.9$, $\beta=5.0$ and $H=0.18$ kpc.

The threshold for unresolved sources is set at $1\times 10^{-9} \; \rm ph \; cm^{-2} \ s^{-1}$. Given this threshold, along with a slope $a$ and $L_{\rm min}$ for the luminosity function, the total number of sources can be determined by requiring that the number of sources above this threshold match what is observed in the {\tt 3FGL} catalog.

We consider as a reference a simulation based on the {\it baseline model} described in the {\tt 3FGL} paper \cite{Acero2015a}, and label it {\it model 1} in \autoref{tab:lum}. For this model, the slope and minimum luminosity are set to the following values: $a=1.8$ and $L_{\rm min} = 2\times10^{34} \; \rm ph \; s^{-1}$. Tuning the number of sources so that the number of the ones above threshold correspond to the catalog yields $\simeq 29000$ sources for this model. 

In addition, we test two different scenarios that provide estimates of the upper and lower limits of our predictions. {\it Model 2} incorporates a harder luminosity function with index $a=1.5$ and a higher minimum luminosity $L_{\rm min}=4 \times 10^{34} \; \rm ph \; s^{-1}$. We find $\simeq 3000$ sources for this model. {\it Model 3} has a softer luminosity function $a=2.1$ and a lower value for $L_{\rm min} = 1 \times 10^{34} \; \rm ph \; s^{-1}$. This yields  $\simeq  210000$ sources. Note that with a supernova rate in our Galaxy of about $2$ per century \cite{Diehl:2006cf}, and given that these sources are active in gamma rays for about 1 Myr, we expect only $\simeq 2 \times 10^4$ of these sources in the Galaxy. {\it Model 3} with $2.1 \times 10^5$ sources can thus be considered as an extreme upper limit, under the assumption that most of these sources result as byproducts of supernova explosions.

\begin{table}
		\centering
		\begin{tabular}{cccc}
			\toprule
				& $a$ & $L_{\rm min} \ [\rm ph \ s^{-1}]$ & $N$ \\
			\midrule
			model 1 &	1.8 & $ 2\times 10^{34}$ & 29000 \\
			\cmidrule(lr){1-4}
			model 2 &	1.5 & $ 4\times 10^{34}$ & 3000 \\
		\cmidrule(lr){1-4}
			model 3 &	2.1 & $ 1\times 10^{34}$ & 210000 \\
 \bottomrule
		\end{tabular}
		\caption{Different scenarios considered in this work for the parameters associated to the gamma-ray source luminosity function.}
		\label{tab:lum} 
\end{table}

In order to estimate statistical fluctuations we produce 100 simulations for each model and report the mean and standard deviation in the following section. 

\subsection{Results: Contribution of unresolved sources at different energies}
\label{sec:UPS_results}

\begin{figure}[h]
\includegraphics[width=0.9\textwidth]{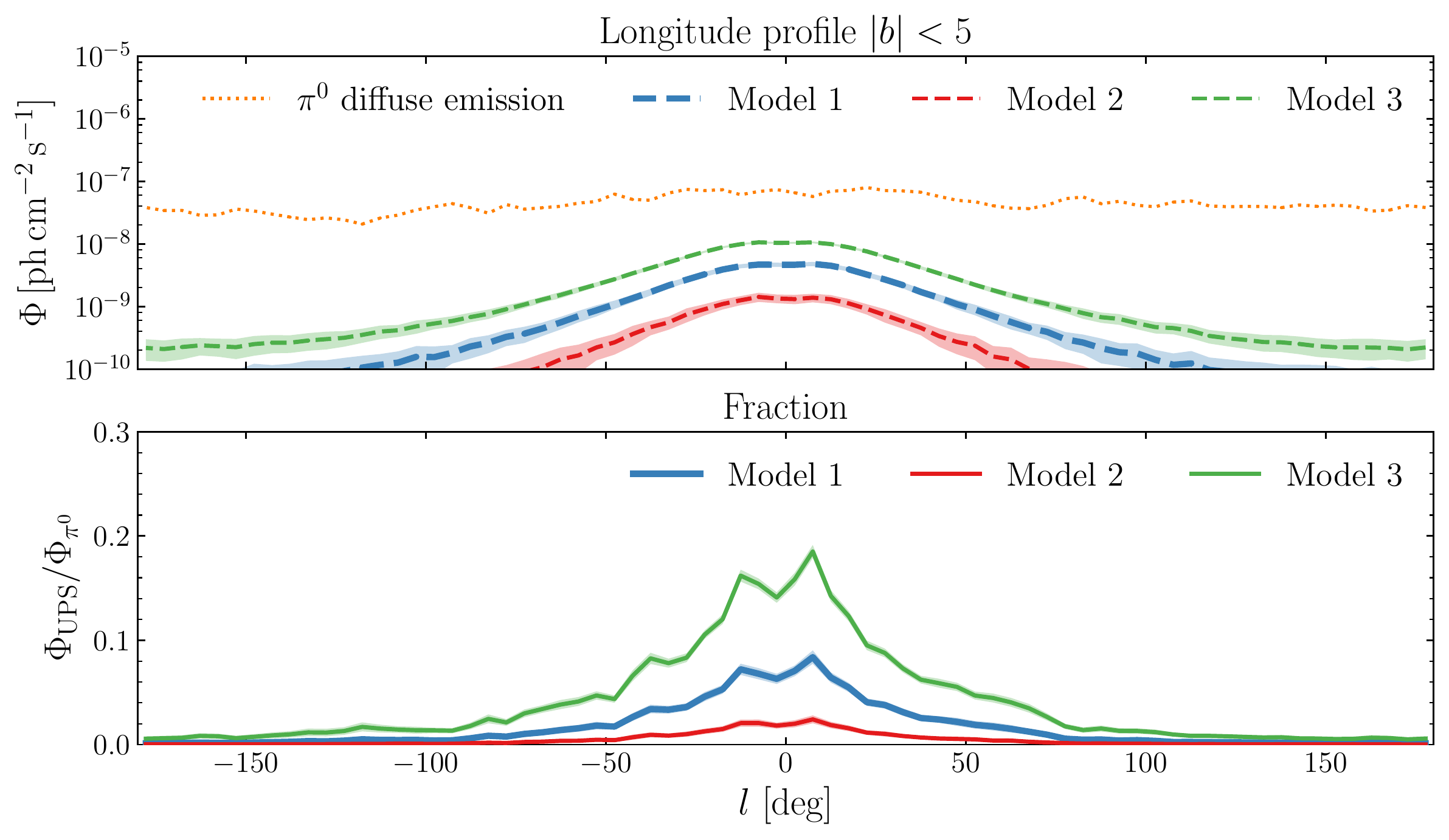}
\caption{{\bf Longitude profiles.}
{\bf Upper panel:} We show the profile of the emission unresolved point sources $\Phi_{\rm UPS}$, and the diffuse hadronic emission $\Phi_{\rm \pi^0}$ from the Galactic plane integrated from $1\text{--}100$ GeV. {\bf Lower panel:} We show the corresponding fraction of unresolved source flux over the diffuse emission. 
The colored bands indicate the estimator of the standard deviation evaluated from 100 simulations.}
\label{fig:longitude}
\end{figure}

\begin{figure}[h]
\includegraphics[width=0.9\textwidth]{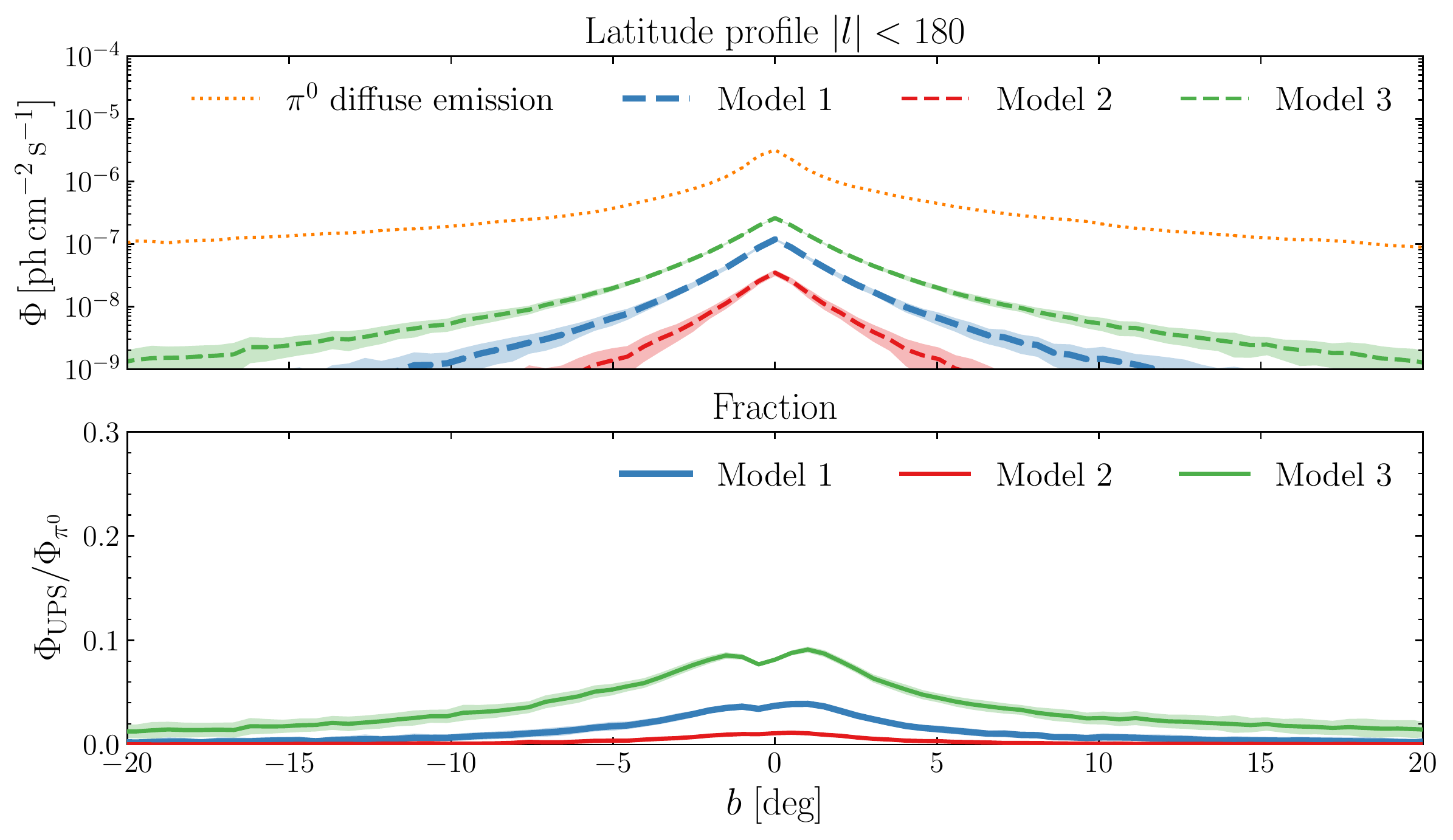}
\caption{{\bf Latitude profiles.}
{\bf Upper panel:} We show the $b$ profile of the emission unresolved point sources $\Phi_{\rm UPS}$, and the diffuse hadronic emission $\Phi_{\rm \pi^0}$  integrated from $1\text{--}100$ GeV and over the whole Galactic plane. {\bf Lower panel:} We show the corresponding fraction of unresolved source flux over the diffuse emission. Colored bands as in the previous plot.}
\label{fig:latitude}
\end{figure}

\begin{figure}[h]
\centering
\includegraphics[width=0.7\textwidth]{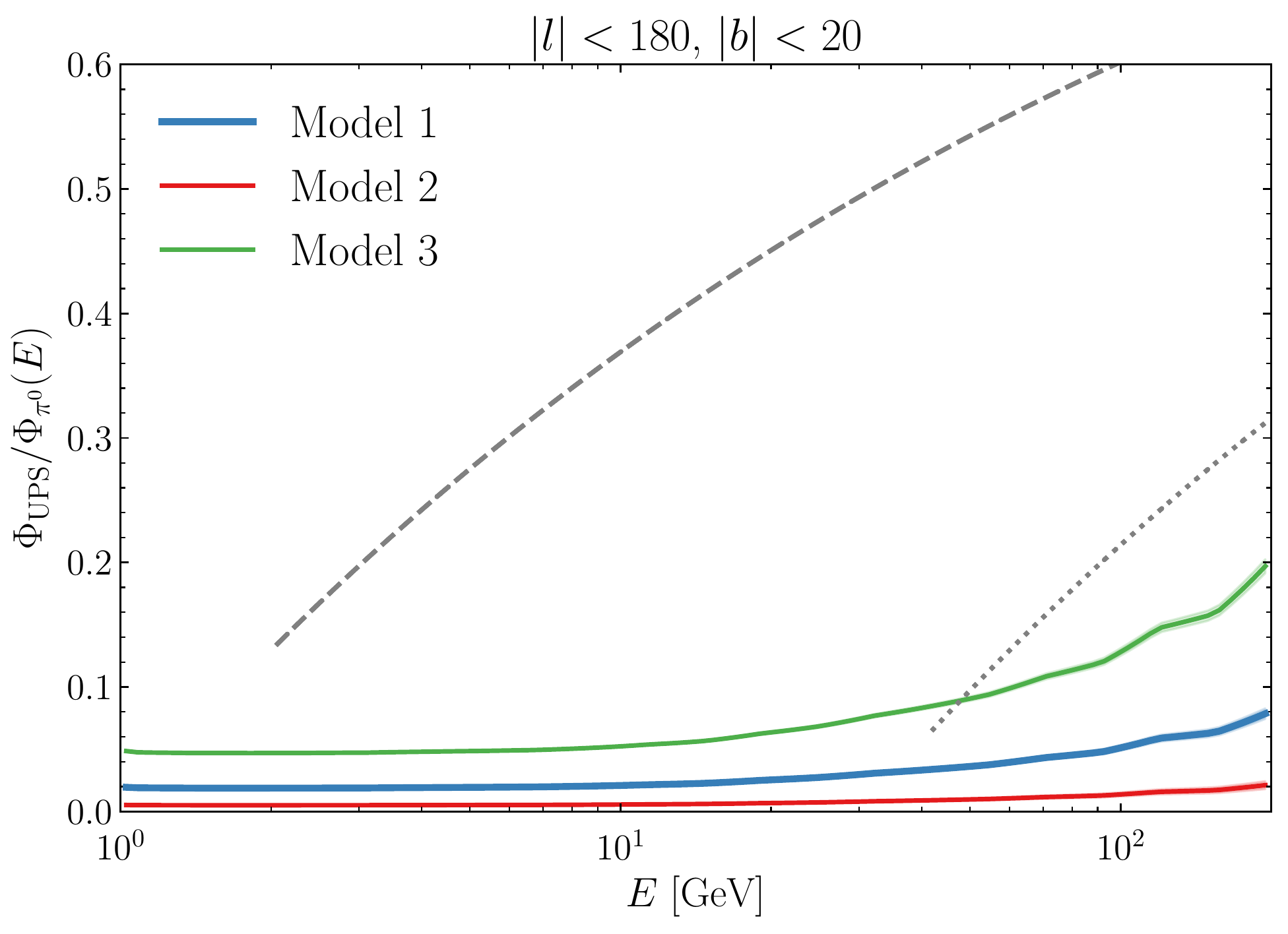}
\caption{{\bf Unresolved sources relative contribution over the whole Galactic plane.} {We show the ratio of the flux from unresolved point sources ($\Phi_{\rm UPS}$) to the hadronic emission} as outlined in \autoref{sec:proton_slope_results}, averaged over all longitudes. 
The dashed line shows the fraction that is needed to harden the total emission from a power-law with index $2.7$ to one with index $2.5$ (normalized at 1 GeV), i.e. \autoref{eq:frac}. 
The dotted line shows the same calculations but normalized at 30 GeV and for a change in index from $2.5$ to $2.3$, and should be interpreted as the fraction of unresolved sources needed to induce the additional hardening at high energies seen in \autoref{fig:index}.}
\label{fig:fraction1}
\end{figure}

\begin{figure}[h]
\centering
\includegraphics[width=0.7\textwidth]{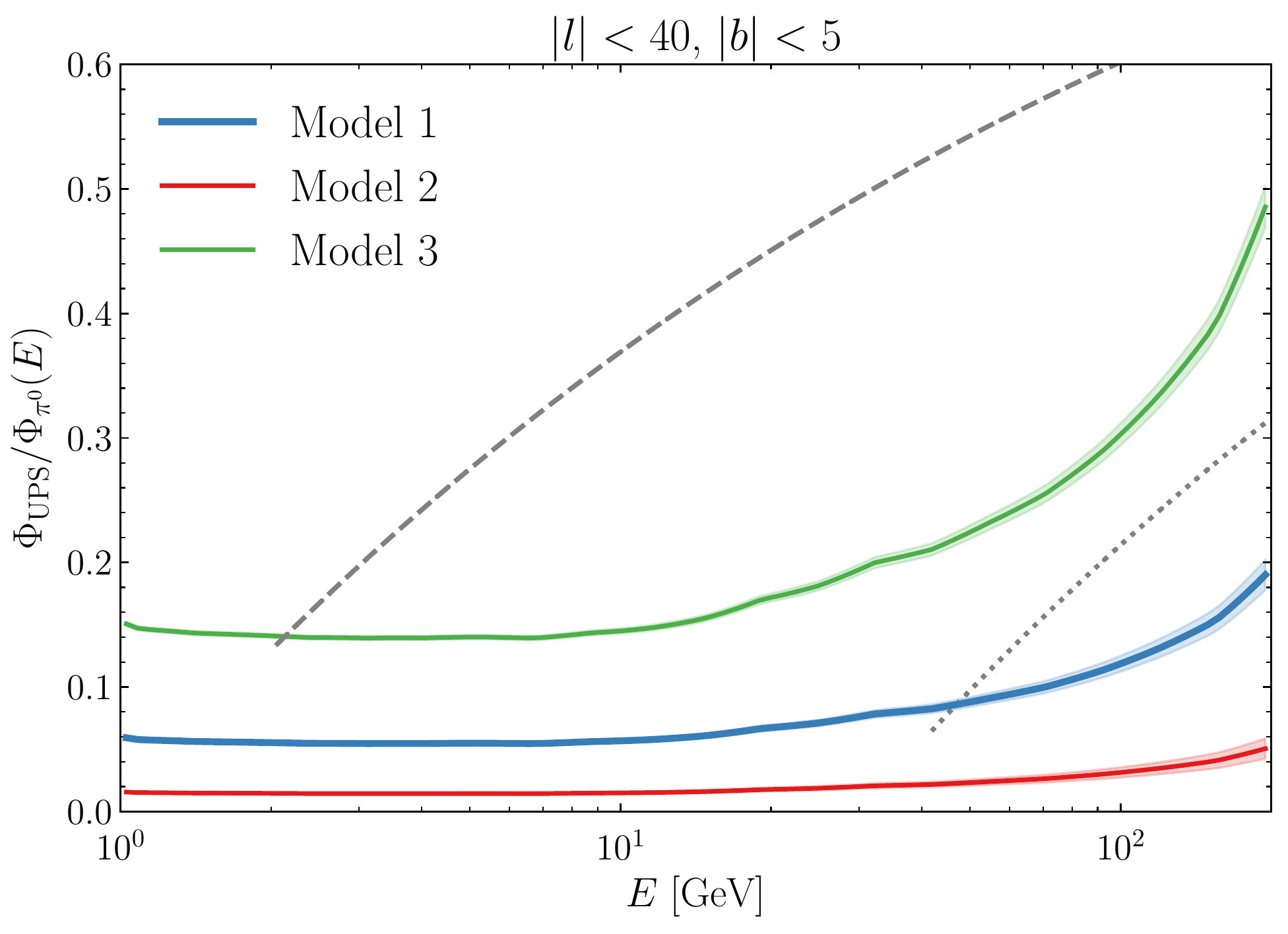}
\caption{{\bf Unresolved sources relative contribution in the inner Galaxy.} We show the ratio between the gamma-ray flux associated to our simulated population of unresolved point sources ($\Phi_{\rm UPS}$) to the diffuse hadronic emission as measured in \autoref{sec:proton_slope_results} averaged over the inner Galactic plane ($|l|<40$ and $|b|<5$): This longitude interval covers the first three radial bins considered in our analysis. The dashed line shows the fraction that is needed to go from a power-law with index 2.7 to 2.5 (with normalization at 1 GeV), i.e. \autoref{eq:frac}. The dotted line shows the same calculations but normalized at 30 GeV and for a change in index from $2.5$ to $2.3$, and should be interpreted as the fraction of unresolved sources needed to induce the additional hardening at high energies seen in \autoref{fig:index}.}
\label{fig:fraction2}
\end{figure}

In \autoref{fig:hists} we show the number of sources per flux interval for the three reference models described above, compared with all the identified Galactic sources from the {\tt 3FGL} catalog (orange circles), and with all the identified sources (including the extra-Galactic ones) with $|b|<5$ (black boxes): All the models considered are compatible with those data. 

In \autoref{fig:longitude} and \autoref{fig:latitude} we show the longitude and latitude profile of the flux from unresolved sources compared to the diffuse $\pi^0$ emission outlined by our analysis (\autoref{sec:proton_slope_results}): From these plots we can see that our {\it model 1} reproduces the few percent contribution of unresolved sources to the diffuse emission found in \cite{Acero2015a}. We remark that the fraction from unresolved sources can be an order of magnitude larger in the inner Galaxy compared to the outer plane. 

\autoref{fig:fraction1} and \autoref{fig:fraction2} show the fraction of the flux associated to unresolved sources to the diffuse pion emission as a function of energy, averaged over the full ROI ($|l|<180$ and $|b|<20$) and the inner Galaxy ($|l|<40$ and $|b|<5$), respectively. The second region in \autoref{fig:fraction2} corresponds to the region where the hardening is particularly pronounced (namely the $R = 4.5\text{--}5.5 \rm \; kpc$ range). In both plots, the dashed line represents the fraction that is needed to mimic the hardening by inducing a change of the power-law associated to the diffuse emission from $2.7$ to $2.5$, with normalization set at 1 GeV, as estimated in \autoref{back-of-the-envolope} (\autoref{eq:frac}). The dotted line is similar, but is normalized at 30 GeV and should be interpreted as the fraction needed to induce the additional hardening from low to high energies present in \autoref{fig:index}.

Over the entire Galactic plane, all the models considered predict fractions well below the  ``warning'' lines in the whole energy range considered (1--100 GeV). In the inner Galaxy, instead, {\it model 3} clearly overshoots the dotted line. However, we emphasize that such a scenario largely overproduces the total number of Galactic gamma-ray point sources sources expected in the Galaxy given our current knowledge on the SN rate, and can to be considered as a rather extreme upper limit on our estimate.  


Given these results, we can conclude that the overall trend outlined with the \skyfact-based analysis cannot be mimicked by a population of unresolved point sources with increasing density towards the inner Galaxy, especially beyond the molecular ring ($R\simeq 4.5$ kpc). The result is substantially unchanged if the tuning is performed on the {\tt 3FHL} catalog.

A useful visualization is provided in \autoref{fig:spec_GR}, where we show the UPS flux associated to {\it model 1} added to the reference power law (compatible with the local slope). For $R > 4.5$ kpc, according to our reference model, the impact of the unresolved sources is clearly not enough to account for the harder gamma-ray emission. However, the UPS component can be responsible for the hardening in the second radial bin. 

Finally, we remark that,  since the 
dotted ``warning'' line associated to the $30$ GeV normalization is actually close to our fiducial model, we can not exclude the possibility that the trend in the high-energy domain (\autoref{fig:index}), which is more pronounced with respect to the overall one,  could partly be due to a contribution from the UPS population in the inner Galaxy.

\section{Summary}

In this paper we have decomposed the gamma-ray emission measured by Fermi-LAT along the Galactic plane ($|b|<20^{\circ}$) in the $0.3\text{--}230$ GeV energy range into different components, including hadronic emission associated to 9 Galactocentric rings, Inverse Compton emission, isotropic diffuse background, point sources and extended sources, and Fermi bubbles. 

The study was based on a penalized maximum likelihood analysis performed with the \skyfact package, which is a hybrid of template fitting and image reconstruction methods and allows the templates to vary in both morphology and spectrum.

We provided evidence for a progressive hardening of the gamma-ray hadronic emission towards the inner Galaxy, which corresponds to a hardening in the CR proton spectrum. 
We showed that this feature is robust with respect to different variations of the analysis setup, and demonstrated that it can be recovered at high energies even if we allow for a break in the power-law fit at 30 GeV. The latter point seems to disfavor an interpretation based on a preeminent role of advection with respect to diffusion in the inner Galaxy due to a significant contribution of streaming instability that lowers the diffusive escape time.

In the second part we estimated the contribution of unresolved point sources via a dedicated Monte Carlo simulation based on our current knowledge of the luminosity function and spatial distribution of Galactic sources (mainly supernova remnants, pulsar wind nebulae and pulsars) and tuned on the {\tt 3FGL} catalog. We considered a fiducial model and extreme scenarios designed to bracket the large uncertainties involved (in particular on the slope and lower limit of the luminosity function). 
The ratio between the flux associated to the simulated population of unresolved sources and the diffuse hadronic emission along the Galactic plane stays below 20\% for all those models up to $\simeq 100$ GeV. We can safely conclude that the progressive hardening under scrutiny is unlikely to be the result of increasing contamination of unresolved point sources in the inner Galaxy for $R > 4.5$ kpc, while further study is needed in the very inner regions of the Galaxy.

\section*{Acknowledgements}

We thank P. Lipari for useful comments and suggestions about the unresolved source population. We acknowledge insights by E. Amato, C. Evoli, D. Grasso, L. Tibaldo, and useful discussions within the CTA consortium. We thank SURFsara (\href{www.surfsara.nl}{www.surfsara.nl}) for the support in using the Lisa Computer Cluster. E.S.~and 
C.W.~acknowledge the support of NWO through a Vidi grant.

\newpage
\bibliographystyle{JHEP}
\bibliography{dragonpaper}


\end{document}